\documentclass{article}

\usepackage[final, nonatbib]{tackling_climate_workshop_style}

\usepackage[utf8]{inputenc} % allow utf-8 input
\usepackage[T1]{fontenc}    % use 8-bit T1 fonts
\usepackage{url}            % simple URL typesetting
\usepackage{booktabs}       % professional-quality tables
\usepackage{amsfonts}       % blackboard math symbols
\usepackage{nicefrac}       % compact symbols for 1/2, etc.
\usepackage{microtype}      % microtypography
\usepackage{xcolor}
\usepackage[dvipsnames]{xcolor}
\definecolor{urlcolor}{RGB}{125, 84, 229} % purple
\definecolor{citationcolor}{RGB}{50, 17, 133} % darkpurple
\definecolor{bostonuniversityred}{rgb}{0.8, 0.0, 0.0}
\usepackage{graphicx}
\usepackage{amsmath}
\usepackage{caption}
\usepackage[backend = bibtex, style = ieee, doi = false,    isbn = false, url = false, eprint = false]{biblatex}
\DeclareFieldFormat{note}{\mkbibbrackets{\url{#1}}}
\usepackage{titlesec}

\usepackage{hyperref}
\hypersetup{
    colorlinks = true, % instead of boxes
    linkcolor = citationcolor, % color for internal links (ToC, equations, etc.)
    citecolor = citationcolor, % citations
    urlcolor = urlcolor % URLs
}

\usepackage[capitalise]{cleveref} % for Appendix numbering

\bibliography{references.bib}

\title{Mass Conservation on Rails -- Rethinking Physics-Informed Learning of Ice Flow Vector Fields}

\author{%
  Kim Bente \\
  School of Computer Science\\
  The University of Sydney\\
  Sydney, Australia\\
\texttt{kim.bente@sydney.edu.au}\\
  \And
  Roman Marchant\\
  Human Technology Institute\\
  University of Technology Sydney\\
  Sydney, Australia\\
  \texttt{roman.marchant@uts.edu.au}\\
  \And
  Fabio Ramos\\
  NVIDIA\\
  Seattle, USA\\
  School of Computer Science\\
  The University of Sydney\\
  Sydney, Australia\\
  \texttt{fabio.ramos@sydney.edu.au}\\
}

\begin{document}

\maketitle
\vspace{-1em}

\begin{abstract}
To reliably project future sea level rise, ice sheet models require inputs that respect physics. Embedding physical principles like mass conservation into models that interpolate Antarctic ice flow vector fields from \textit{sparse \& noisy} measurements not only promotes physical adherence but can also improve accuracy and robustness. While physics-informed neural networks (PINNs) impose physics as soft penalties, offering flexibility but no physical guarantees, we instead propose divergence-free neural networks (dfNNs), which enforce local mass conservation exactly via a vector calculus trick. Our comparison of dfNNs, PINNs, and unconstrained NNs on ice flux interpolation over Byrd Glacier suggests that "mass conservation on rails" yields more reliable estimates, and that \textit{directional guidance}, a learning strategy leveraging continent-wide satellite velocity data, boosts performance across models.
\end{abstract}
% improve text distribution & breaks
\vspace{-1em}

\section{Introduction \& Background}

By integrating physical laws into data-driven learning, \textbf{physics-informed machine learning} (PIML) \cite{karniadakis_physics-informed_2021} has reshaped ML's impact across the physical sciences, including notable advances in climate science \cite{reichstein_deep_2019, rolnick_tackling_2022, monteleoni_climateinformatics_bookchapter_2013}. Many of the Essential Climate Variables (ECVs) \cite{world_meteorological_organization_essential_2024}, such as ocean currents, groundwater, and glacier flow, take the form of spatial vector fields, $\mathbf{v}(x, y)$, that describe Earth’s fluid dynamics. These flows are governed by the continuity equation, which ensures that mass of the flowing fluid is conserved \cite{kundu_chapter_2016}. For steady, incompressible flow, this mass conservation constraint reduces to the \textit{divergence-free condition}, $\nabla \cdot \mathbf{v} = 0$, enforcing that inflow equals outflow at any point in space. Diverse models of groundwater \cite{zhang_new_2024}, ocean eddies and currents \cite{berlinghieri_gaussian_2023, sosanya_dissipative_2022}, tidal flows \cite{vennell_divergence-free_2009}, ice sheets \cite{morlighem_mass_2011, teisberg_machine_2021}, glaciers \cite{steidl_physics-aware_2024}, ocean surface winds \cite{fan_modeling_2016}, and large-scale atmospheric circulations \cite{narcowich_divergence-free_2007, loptien_global-scale_2018} have incorporated the \textbf{divergence-free condition} to promote physically consistent behaviour, thereby improving accuracy and robustness, which facilitates reliable downstream use.

Here, we focus on an application with especially grave implications\,---\,modelling how ice flows across the Antarctic Ice Sheet (AIS). Holding  the ice equivalent of \textasciitilde58 m global mean sea level rise \cite{morlighem_deep_2020}, the AIS is the largest potential contributor to rising seas \cite{ipcc_summary_2022}. By 2100, sea level rise is projected to impose annual flood damage costs of \textasciitilde 2\% of GDP \cite{jevrejeva_flood_2018} and to displace \textasciitilde 360 M people living on flood-prone land under RCP~4.5 \cite{kulp_new_2019}. However, extreme polar conditions and remoteness severely limit data collection, making AIS modelling challenging, yet scientifically paramount. Studies have found that unconstrained interpolations of sparse and noisy ice thickness measurements from airborne radar surveys produce unrealistic behaviour in numerical ice sheet models due to flux divergences \cite{seroussi_ice_2011, morlighem_mass_2011} (\textit{flux}: transport of ice volume per unit width and time). In response to this issue, numerical inversions \cite{morlighem_mass_2011, morlighem_deep_2020} and, more recently, ML approaches \cite{teisberg_machine_2021, steidl_physics-aware_2024, liu_physics-informed_2024} have been developed to \textbf{reduce flux artefacts by constraining ice flux interpolations with mass conservation}.\\
\begin{figure}[t!]  % h = here, t = top, b = bottom, p = page
    \centering
    \includegraphics[width=\textwidth]{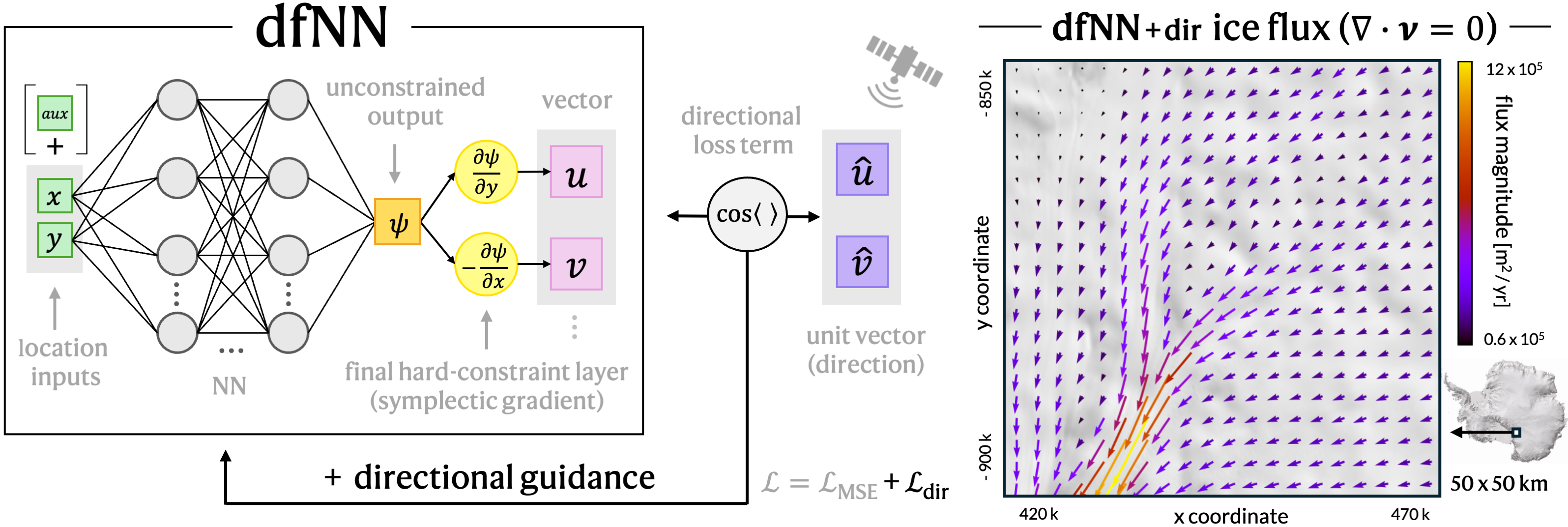}
    \caption{Diagram of the divergence-free NN (dfNN) model architecture and \textit{directional guidance} learning strategy (left). The quiver plot (right) shows an on-grid ice flux reconstruction by the dfNN with \textit{directional guidance} (best model) for a subset of the experimental region over Byrd Glacier.}
    \label{fig:intro_model}
    % improve text distribution & breaks
    \vspace{-0.2cm}
\end{figure}

Both proposed ML approaches \cite{teisberg_machine_2021, steidl_physics-aware_2024} employ physics-informed neural networks (PINNs) \cite{raissi_physics_2017}, a widely used PIML framework. PINNs incorporate an additional loss term that quantifies violations of physical principles, to learn not only to fit the training data, but to simultaneously minimise physical inconsistencies, such as flux artefacts. Although PINNs have the flexibility to incorporate multiple physical principles at once, as demonstrated in \cite{teisberg_machine_2021} and \cite{steidl_physics-aware_2024}, and are hence broadly adopted, the penalty-based \textit{`soft'} constraining does not guarantee physical consistency, which has been found to hinder generalisation performance in the face of data sparsity and noise \cite{greydanus_hamiltonian_2019, macedo_learning_2010}. Furthermore, unstable trade-offs between data fit and physics can also affect convergence \cite{richter-powell_neural_2022}. In this work, we show how lesser known hard-constrained \textbf{divergence-free NNs} (dfNNs, see \cref{fig:intro_model}), rooted in  \cite{kuroe_learning_1998, greydanus_hamiltonian_2019, richter-powell_neural_2022}, can be used to \textbf{model exactly mass-conserving ice flux vector fields}. To assess competing PIML paradigms for modelling divergence-free vector fields, we compare hard-, soft-, and unconstrained NN models\,---\,dfNNs, PINNs, and regular NNs\,---\,on ice flux reconstruction for Byrd Glacier, Antarctica, shown in \cref{fig:intro_model}, evaluating both predictive accuracy and physics adherence. Lastly, informed by the application context \cite{rolnick_position_2024}, we test two extensions across all three models: (i) \textit{directional guidance}, a learning strategy that leverages continent-wide ice velocity observations from satellites via a directional loss term $\mathcal{L}_{\text{\tiny dir}}$ to align predicted flow with this directional information beyond sparse flux observation locations, and (ii) incorporating auxiliary predictors (e.g., surface elevation). Reproducible experiments and implementations in PyTorch \cite{paszke_pytorch_2019} are available at \url{https://github.com/kimbente/mass_conservation_on_rails}.

\section{Exact mass conservation with dfNNs}

% Modelling 
% divergence-free condition
\textbf{\textit{Introducing dfNNs.}} In this work, we address the problem of learning 2D vector fields with NNs under the constraint of local mass conservation. Let \( \mathbf{v}:\mathbb{R}^2 \to \mathbb{R}^2 \) denote a 2D vector field, expressed as \( \mathbf{v}(x, y) = (u(x, y),\, v(x, y)) \), where \( u \) and \( v \) are the vector components in \( x \)- and \( y \)-direction, respectively. The divergence of a vector field is defined as the sum of the partial derivatives of its components, $\nabla \cdot \mathbf{v} = \partial u/\partial x + \partial v/\partial y$, which quantifies the local rate of expansion or compression. In balanced flows, this quantity must equal zero. The first NN model to integrate the divergence-free constraint into the model architecture was introduced in early work of Kuroe et~al.~\cite{kuroe_learning_1998}. The proposed approach, \textit{model inclusive learning}, leverages the property that the symplectic gradient of a scalar stream function is by design divergence-free (see \cite{kundu_chapter_2016} for vector calculus background). By training the NN to predict the stream function rather than the vector components, akin to the change of variables trick, and attaining the vector components by taking the symplectic gradient in a deterministic differential step, the network outputs continuous, exactly divergence-free vector fields, i.e. with $\nabla \cdot \mathbf{v} = 0$. Closely related model architectures have reappeared more recently: Hamiltonian Neural Networks \cite{greydanus_hamiltonian_2019}, which were introduced with energy rather than mass conservation in mind, and Neural Conservation Laws \cite{richter-powell_neural_2022}, a theoretically comprehensive framework that extends to solving general continuity equations, both impose the divergence-free constraint through a construction equivalent to \cite{kuroe_learning_1998}. Since the original model \cite{kuroe_learning_1998}, like \cite{sosanya_dissipative_2022}, addresses vector fields beyond the purely divergence-free case by decomposing them into curl-free and divergence-free components, we refer to the divergence-free component as a divergence-free neural network (dfNN) throughout this paper. The left panel of \cref{fig:intro_model} illustrates the dfNN architecture. Spatial coordinates $(x, y)$, together with optional auxiliary inputs, are passed through a feed-forward neural network whose width and depth can be adjusted to the task complexity. For each input location, the network outputs a scalar \textit{stream function} $\psi$ (refer to \cite{kundu_chapter_2016}). As proposed by \cite{kuroe_learning_1998} this unconstrained scalar field is then mapped to a strictly divergence-free vector field by applying the symplectic operator. Its components, given by the symplectic gradient $(\partial \psi/\partial y, -\partial \psi/\partial x)$, define the velocity field $(u, v)$ and can be computed efficiently via automatic differentiation in ML frameworks such as PyTorch \cite{paszke_pytorch_2019}, all in a fully mesh-free manner. 

\textbf{\textit{Modelling ice flux as a divergence-free flow.}}  \textit{Ice flux} denotes the horizontal transport of ice volume per unit width and time, with volume equivalent to mass under incompressibility. Ice flux vectors $\mathbf{v}$ have units
$[\mathrm{m}^3\,\mathrm{m}^{-1}\,\mathrm{yr}^{-1}]$, or simply $[\mathrm{m}^2\,\mathrm{yr}^{-1}]$, and are defined as the product $\mathbf{v} = h \cdot \mathbf{s}$ where ice thickness $h \in \mathbb{R}$ is in $[\mathrm{m}]$ and horizontal velocity $\mathbf{s} \in \mathbb{R}^2$ is in $[\mathrm{m}\,\mathrm{yr}^{-1}]$. Following \cite{morlighem_mass_2011} and \cite{teisberg_machine_2021}, we take the measured surface-level velocity as a proxy for depth-averaged velocity, to model the depth-integrated transport of ice. Given the AIS's quasi-steady flow, minimal thickness changes, and a net surface/basal mass balance that is negligible relative to fluxes along ice streams \cite{morlighem_deep_2020, morlighem_mass_2011}, we model the problem as a spatial (time-independent) process under divergence-free flow assumptions. As motivated above, the resulting spatial ice flux fields are a critical input for numerical ice sheet models \cite{seroussi_ice_2011, morlighem_mass_2011}, and, conversely, ice thickness maps can be extracted with $h = \mathbf{v} / \mathbf{s}$ \cite{morlighem_mass_2011, teisberg_machine_2021, steidl_physics-aware_2024}.

\textbf{\textit{Directional guidance.}} While observations of environmental processes are typically sparse, InSAR satellites provide continent-wide ice surface velocity observations $\mathbf{s}$ on a dense grid over Antarctica \cite{mouginot_measures_2019}. Given $\mathbf{v} = h \cdot \mathbf{s}$, these observations only determine the direction of $\mathbf{v}$, but not its magnitude. Still, this partial information can help improve and further constrain reconstructions in unsurveyed regions. In a learning strategy that we refer to as \textit{directional guidance}, conceptually similar to the physical guidance term in PINNs \cite{raissi_physics-informed_2019}, we integrate a loss term $\mathcal{L}_{\text{dir}}$ that quantifies the directional misalignment between domain-wide unit velocity observations $\hat{\mathbf{s}}$, and corresponding normalised model-predicted vectors $\hat{\mathbf{v}}$, via the cosine similarity, as defined in \cref{eq:directional_loss},
\begin{equation}
\mathcal{L} = (1 - w_{\text{dir}})\cdot\mathcal{L}_{\text{MSE}} + w_{\text{dir}} \cdot\mathcal{L}_{\text{dir}}, 
\quad \text{with} \,\ \mathcal{L}_{\text{dir}} = 1 - \cos\!\left(\hat{\mathbf{s}},\,  \hat{\mathbf{v}}\right) \, \in [0,2].
\label{eq:directional_loss}
\end{equation}
Weighted by $w_{\text{dir}}$, $\mathcal{L}_{\text{dir}}$ is integrated with the regular training loss $\mathcal{L}_{\text{MSE}}$, so that gradient updates simultaneously fit flux observations over training regions and align predicted directions with satellite‐derived flow directions elsewhere.

\section{Experiments \& Results}

\textbf{\textit{Data}.} To assess how best to enforce local mass conservation constraints in learning vector fields, we compare hard-constrained dfNNs \cite{kuroe_learning_1998, greydanus_hamiltonian_2019, richter-powell_neural_2022}, soft-constrained PINNs \cite{raissi_physics_2017}, and unconstrained NNs using real Antarctic observations. Similar to \cite{teisberg_machine_2021}, we focus on a $200 \times 200$ km region over Byrd Glacier (see \cref{fig:intro_model}), a fast-flowing outlet glacier with sufficiently dense observations to withhold data for testing. Ice flux observations $\mathbf{v}$ are obtained by combining ice thickness measurements $h$ from the comprehensive \textit{Bedmap} data collection \cite{fremand_antarctic_2023} with ice surface velocities $\mathbf{s}$ from NASA's \textit{MEaSUREs Phase-Based Antarctica Ice Velocity Map} \cite{mouginot_measures_2019}. The data are split into train and test regions using a chequerboard pattern with 15 km squares (\cref{fig:checkerboard}), yielding 27,172 training and 22,045 testing points. Further details on data sources and preprocessing are provided in \cref{app:data}.

\textbf{\textit{Experiments}.} In total, we train nine models and evaluate their reconstruction performance on the test set. These include the three base models, their \textit{directionally guided} counterparts, and variants incorporating surface elevation from \textit{Bedmap} as an auxiliary input. All models share a consistent NN backbone and are trained with PyTorch’s AdamW optimiser with weight decay, which we found to improve convergence. We evaluate predictive accuracy using RMSE and MAE, and quantify physics violations with the Mean Absolute Divergence (MAD, defined in \cref{app:mad}). We also monitor the carbon emissions of our experiments with \textit{CodeCarbon}, and report them in \cref{app:emissions}. Further training details are provided in \cref{app:training}.
\begin{figure}[t]  % h = here, t = top, b = bottom, p = page
    \centering
    \includegraphics[width = \textwidth]{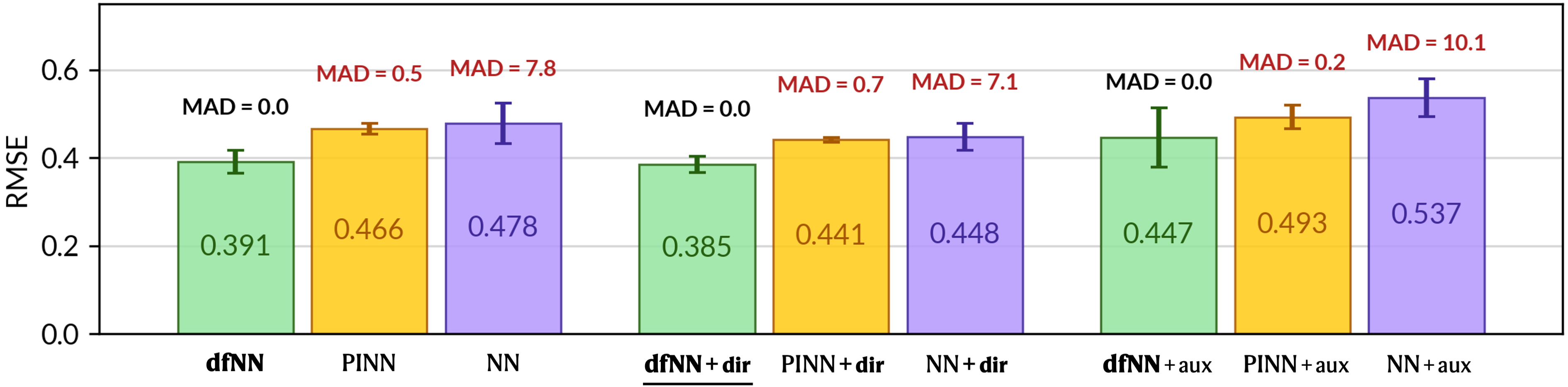}
    \caption{Test RMSE ($\downarrow$) comparison across all model variants, averaged over five runs. Boxed values indicate mean RMSE, with error bars showing $\pm$ std. MAD (top) denotes the Mean Absolute Divergence. \textbf{dfNNs} (proposed, in bold) outperform PINNs \& NNs, while \textbf{\textit{directional guidance}} (proposed, in bold) improves all models and yields the best-performing variant, dfNN + dir (underlined).}
    \label{fig:rmse_barplot}
    % improve text distribution & breaks
    \vspace{-0.2cm}
\end{figure}

\begin{figure}[h]  % h = here, t = top, b = bottom, p = page
    \centering
    % order: graphic, caption, label
    \includegraphics[width = \textwidth]{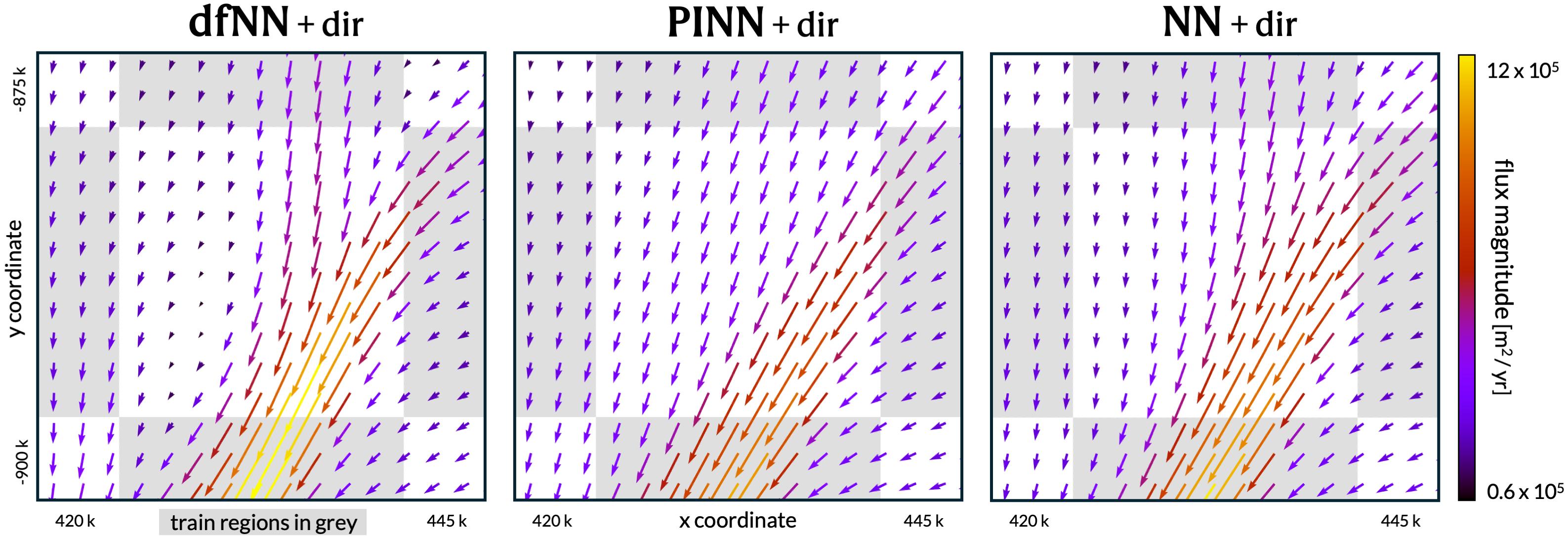}
    \caption{Gridded predictions of models + dir (best variant per model) for a small test region (white).}
    \label{fig:preds}
\end{figure}

\textbf{\textit{Results}.} Test results from all nine models, averaged over five independent training runs, are visualised in \cref{fig:rmse_barplot}, with full metrics (mean ± std) reported in \cref{tab:results} (Appendix). Across all metrics, dfNNs (and their variants) consistently yield more reliable reconstructions of unseen ice flux vectors than both PINNs and NNs, underscoring the benefits of enforcing exact mass conservation for both accuracy and physical compliance. PINNs substantially reduce MAD compared to NNs, but their improvement in interpolation accuracy is modest. Furthermore, they do not converge to fully divergence-free solutions (MAD > 0). \textit{Directional guidance} improves performance across all models, with larger relative gains for PINNs and NNs, which have weaker inductive biases than dfNNs. Predictions by all models + dir, the best variant of each model family, are shown in \cref{fig:preds}. By contrast, incorporating surface elevation as an auxiliary predictor degrades performance, with exploratory tests using surface gradients showing an even stronger decline. Overall, the dfNN with \textit{directional guidance} is the best-performing model, producing guaranteed divergence-free interpolations with MAD = 0.
% improve text distribution & breaks
\vspace{-10pt}
\section{Conclusion}
Our experiments on real Antarctic ice flux data highlight the advantages of enforcing known physical inductive biases with hard constraints (dfNNs) rather than loss terms (PINNs). Nonetheless, additional loss term guidance can help hard-, soft-, and unconstrained models alike: Our proposed \textit{directional guidance} learning strategy shows an effective way of leveraging dense satellite data to constrain the direction of environmental flows, boosting performance across all models. Furthermore, the ablations substantiate the premise that parsimonious models generalise best: Adding auxiliary surface predictors like elevation introduces more noise than signal, indicating that relevant topographic flow controls are not expressed at the surface over Byrd Glacier. Overall, our work suggests that when available, hard-constrained "models on rails" should be the method of choice for physics-informed learning, as they offer both physical fidelity and maintain model design flexibility to incorporate additional guidance, while remaining robust under the \textit{noisy \& sparse} data typical of climate and environmental applications in the real world.

% KB: These are hidden in the submission file
\begin{ack}
This research was supported by an Australian Government Research Training Program (RTP) Scholarship, as well as through the Australian Research Council’s Industrial Transformation Training Centre in Data Analytics for Resources and Environments (DARE) (project ID IC190100031).
\end{ack}

%%% REFERENCES %%%

% KB: reduces font size for references to 9 pt. which is allowed
% NOTE: Needs to be wrapped!! Otherwise the appendix fontsize is smaller
{\small
\printbibliography
}
%%%%%%%%%%%%%%%%%%%%%%%%%%%%%%%%%%%%%%%%%%%%%%%%%%%%%%%%%%%%

% KB: switch back to normnal font size

% Start on new page
\newpage

% switches to A, B, C 
\appendix
% KB: Assure "Appendix before every sections
% \renewcommand\thesection{Appendix \Alph{section} \, --}

\titleformat{\section} % redefine section headings
  {\normalfont\large\bfseries}  % formatting
  {Appendix \Alph{section} \,  --}  % label
  {0.6em}                       % spacing between label and title
  {}  

\section{Results Table}
\label{app:table}

% \normalfont
\cref{tab:results} contains all experimental results. A visual comparison of RMSE (Root Mean Square Error) and MAD (Mean Absolute Divergence) values can be found above in \cref{fig:rmse_barplot}.

\begin{table}[h]
\caption{Test metrics (mean {\footnotesize ± std}) over 5 runs. MAD denotes the Mean Absolute Divergence. Lower values $\downarrow$ indicate better performance. The overall best model is shown in bold, and physics violations (flux divergences) are highlighted in red. 
The model order follows \cref{fig:rmse_barplot}.}
\label{tab:results}
\centering
\begin{tabular}{lccc}
\toprule
Model (+ extension) &
\multicolumn{1}{c}{\textbf{RMSE $\pm$ std}} &
\multicolumn{1}{c}{\textbf{MAE $\pm$ std}} &
\multicolumn{1}{c}{\textbf{MAD $\pm$ std}}\\
\midrule
dfNN             & 0.391 \footnotesize{± 0.03} & 0.199 \footnotesize{± 0.01} & 0.000 \footnotesize{± 0.00}\\
\textbf{dfNN + dir} & 0.385 \footnotesize{± 0.02} & 0.193 \footnotesize{± 0.01} & 0.000 \footnotesize{± 0.00}\\
dfNN + aux       & 0.447 \footnotesize{± 0.07} & 0.209 \footnotesize{± 0.01} & 0.000 \footnotesize{± 0.00}\\
\midrule
PINN             & 0.466 \footnotesize{± 0.01} & 0.236 \footnotesize{± 0.01} & \textcolor{bostonuniversityred}{0.471 \footnotesize{± 0.06}} \\
PINN + dir       & 0.441 \footnotesize{± 0.01} & 0.221 \footnotesize{± 0.01} & \textcolor{bostonuniversityred}{0.685 \footnotesize{± 0.11}} \\
PINN + aux       & 0.493 \footnotesize{± 0.03} & 0.237 \footnotesize{± 0.01} & \textcolor{bostonuniversityred}{0.235 \footnotesize{± 0.03}} \\
\midrule
NN               & 0.478 \footnotesize{± 0.05} & 0.215 \footnotesize{± 0.01} & \textcolor{bostonuniversityred}{7.792 \footnotesize{± 0.43}} \\
NN + dir         & 0.448 \footnotesize{± 0.03} & 0.205 \footnotesize{± 0.01} & \textcolor{bostonuniversityred}{7.137 \footnotesize{± 0.31}} \\
NN + aux         & 0.537 \footnotesize{± 0.04} & 0.237 \footnotesize{± 0.02} & \textcolor{bostonuniversityred}{10.13 \footnotesize{± 0.75}} \\
\bottomrule
\end{tabular}
\end{table}

\section{Training details}
\label{app:training}

\cref{tab:hypers} contains all hyperparameters used for training, where the reported values correspond to the best-performing settings identified in our investigation. PINNs are trained with an additional divergence reduction step at every epoch, to ensure that the model also learns to reduce divergences over test regions. 

% Added NN params to table
% \usepackage{booktabs}
\begin{table}[h]
\caption{Training hyperparameters used in experiments. Please also refer to the repository's \texttt{configs.py} file on \url{https://github.com/kimbente/mass_conservation_on_rails} directly.}
\label{tab:hypers}
\centering
\small
\begin{tabular}{ll}
\toprule
Hyperparameter & Value \\
\midrule
dfNN learning rate (all model variants) & 0.0001 \\
PINN learning rate (all model variants) & 0.0001 \\
NN learning rate (all model variants) & 0.0005 \\
\midrule
Number of runs & 5\\
Maximum number of epochs & 3000\\
Patience for early stopping & 100 \\
Batch size & 1024\\
\midrule
Number of hidden layers (for all NNs) & 4 \\
Width of each hidden layer (for all NNs) & 64 \\
\midrule
Optimiser & AdamW\\
Weight decay & 0.001 \\
\midrule
Weight $w_{\text{dir}}$ for $\mathcal{L}_{\text{dir}}$ (dfNN + dir, NN + dir)& 0.4\\
\midrule
PINN number of points used per epoch for additional domain divergence reduction step &10,240\\
PINN weight $w_{\text{div}}$ for $\mathcal{L}_{\text{div}}$ (PINN, PINN + aux) & 0.2\\
\midrule
for PINN + dir both $w_{\text{dir}}$ and $w_{\text{div}}$ are halved to avoid underweighing the training loss $\mathcal{L}_{\text{MSE}}$ & \\
\quad Adjusted ($\times$ 0.5) weight $w_{\text{dir}}$ for $\mathcal{L}_{\text{dir}}$ (PINN + dir) & 0.2 \\
\quad Adjusted ($\times$ 0.5) PINN weight $w_{\text{div}}$ for $\mathcal{L}_{\text{div}}$ (PINN + dir)& 0.1\\
\bottomrule
\end{tabular}
\end{table}

\newpage
\section{Mean Absolute Divergence (MAD)}
\label{app:mad}

We introduce the metric Mean Absolute Divergence (MAD) to quantify flux divergences (i.e. violations of local mass conservation). Taking the absolute divergence ensures that local positive and negative divergences, which both violate the divergence-free constraint, do not cancel out. MAD is calculated as defined in \cref{eq:mad}.

\begin{equation}
\text{MAD} \;=\; \frac{1}{N} \sum_{i=1}^{N} 
\left| \nabla \cdot \mathbf{q}(\mathbf{x}_i) \right|
\;=\; \frac{1}{N} \sum_{i=1}^{N} 
\left| \frac{\partial u}{\partial x}(\mathbf{x}_i) + \frac{\partial v}{\partial y}(\mathbf{x}_i) \right|
\label{eq:mad}
\end{equation}

\section{Training and testing data}
\label{app:data}

Byrd Glacier is a major Antarctic outlet glacier draining a basin that holds 6 m sea level equivalent (SLE) \cite{morlighem_deep_2020}. The $200 \times 200$ km region, about the size of Denmark, is characterised by fast, topographically steered flow, and is relatively densely surveyed, allowing us to withhold a subset of data for testing. 

During preprocessing, we apply a uniform firn correction to ice thickness measurements, downsample along densely spaced airborne survey flight lines, and interpolate the velocity grid to the ice thickness point locations to obtain $\mathbf{v}$.

\begin{figure}[h]  % h = here, t = top, b = bottom, p = page
    \centering
    \includegraphics[width = 0.8 \textwidth]{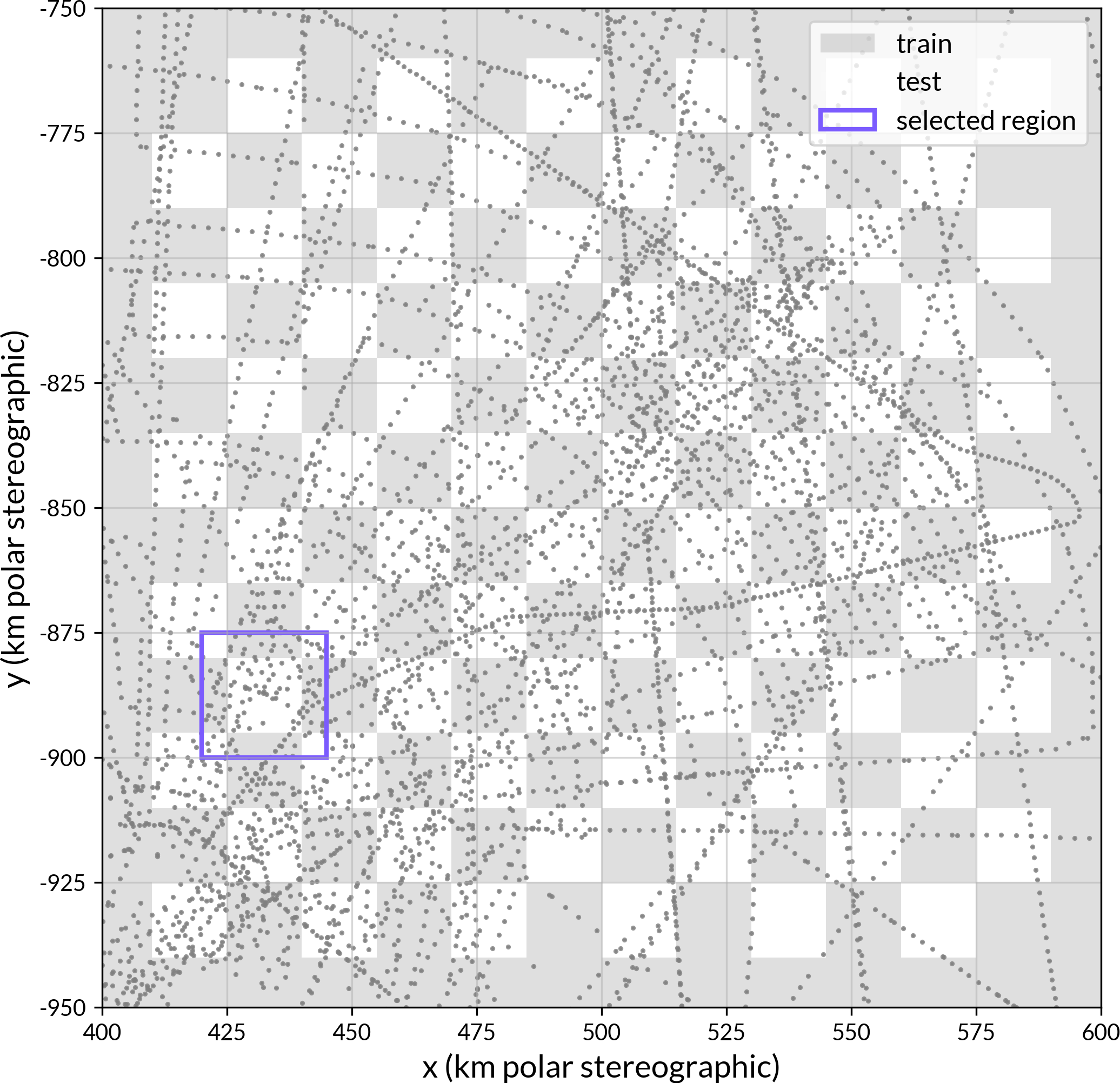}
    \caption{Train-Test chequerboard over Byrd Glacier, Antarctica. Training regions are shown in grey and test regions are shown in white. Points indicate locations with ice flux observations, collected mainly through airborne geophysical surveys, highlighting the anisotropic nature of the `flight line' data. The Antarctic Polar stereographic coordinate system is used (ESPG:3031, \url{https://epsg.io/3031}). The selected region (purple frame) corresponds to the region shown in \cref{fig:preds}.}
    \label{fig:checkerboard}
\end{figure}

\cref{fig:checkerboard} shows the chequerboard pattern which we use to divide the Byrd Glacier domain into training and testing regions.

% Finish with chequerboard

\newpage
\section{Emissions}
\label{app:emissions}
We use \textit{CodeCarbon} to track our experiments' emissions and report them in \cref{tab:comp}.

\begin{table}[h]
\caption{Total computational demands of 5 experiment runs (model training and inference), as tracked by \textit{CodeCarbon}, see \url{https://codecarbon.io/}. (GPU: 1 × NVIDIA RTX 4090, RAM: 63 GB).}
\label{tab:comp}
\centering
\begin{tabular}{lccc}
\toprule
Model ({\footnotesize + extension}) &
\multicolumn{1}{c}{Emissions} &
\multicolumn{1}{c}{Energy consumed} &
\multicolumn{1}{c}{Wall clock time} \\
&
\multicolumn{1}{c}{[in kgCO\textsubscript{2}eq]} &
\multicolumn{1}{c}{[in kWh]} &
\multicolumn{1}{c}{[in minutes]}\\
\midrule
dfNN             & 0.219 & 0.399 & 139.19\\
dfNN + dir       & 0.171  & 0.312 & 171.53\\
dfNN + aux       & 0.236 & 0.430 & 131.29\\
\midrule
PINN             & 0.078 & 0.143 & 62.49\\
PINN + dir       & 0.072 & 0.132 & 75.27\\
PINN + aux       & 0.098 & 0.178 & 102.85\\
\midrule
NN               & 0.076 & 0.138 & 81.89\\
NN + dir         & 0.083 & 0.151 & 90.16\\
NN + aux         & 0.073 & 0.134 & 79.34\\
\bottomrule
\end{tabular}
\end{table}

\end{document}